\documentclass[twoside,11pt]{article}

\usepackage[accepted]{melba}

\usepackage{amsmath,amsfonts}

\melbaid{2023:006}  
\melbaauthors{Patel, Tudiosu, Pinaya, Cook, Goh, Ourselin, Cardoso}
\volume{2}
\firstpageno{172}  
\year{2023}  
\datesubmitted{11/2022}  
\datepublished{04/2023}  

\ShortHeadings{Multi-modal Transformers for Anomaly Detection}{Patel, Tudiosu, Pinaya, Cook, Goh, Ourselin, Cardoso}

\title{Cross Attention Transformers for Multi-modal Unsupervised Whole-Body PET Anomaly Detection}

\author{\name Ashay Patel \orcid{0000-0003-4212-2578} \email ashay.patel@kcl.ac.uk 
	\AND
	\name Petru-Danial Tudiosu \orcid{0000-0001-6435-5079} \email petru.tudosiu@kcl.ac.uk
 	\AND
	\name Walter H.L. Pinaya \orcid{0000-0003-3739-1087} \email walter.diaz\_sanz@kcl.ac.uk
  	\AND
	\name Gary Cook \orcid{0000-0002-8732-8134} \email gary.cook@kcl.ac.uk
  	\AND
	\name Vicky Goh \orcid{0000-0002-2321-8091} \email vicky.goh@kcl.ac.uk
  	\AND
	\name Sebastien Ourselin \orcid{0000-0002-5694-5340} \email sebastien.ourselin@kcl.ac.uk
   	\AND
	\name M. Jorge Cardoso \orcid{0000-0003-1284-2558} \email m.jorge.cardoso@kcl.ac.uk \\
	\addr King's College London, Becket House, 1 Lambeth Palace Rd, London SE1 7EU
}

\begin{document}

\maketitle

\vspace{-1em}
\begin{abstract}
	Cancer is a highly heterogeneous condition that can occur almost anywhere in the human body. \textsuperscript{18}F-fluorodeoxyglucose (\textsuperscript{18}F-FDG PET) is an imaging modality commonly used to detect cancer due to its high sensitivity and clear visualisation of the pattern of metabolic activity. Nonetheless, as cancer is highly heterogeneous, it is challenging to train general-purpose discriminative cancer detection models, with data availability and disease complexity often cited as a limiting factor. Unsupervised learning methods, more specifically anomaly detection models, have been suggested as a putative solution. These models learn a healthy representation of tissue and detect cancer by predicting deviations from the healthy norm, which requires models capable of accurately learning long-range interactions between organs, their imaging patterns, and other abstract features with high levels of expressivity. Such characteristics are suitably satisfied by transformers, which have been shown to generate state-of-the-art results in unsupervised anomaly detection by training on normal data. This work expands upon such approaches by introducing multi-modal conditioning of the transformer via cross-attention i.e. supplying anatomical reference information from paired CT images to aid the PET anomaly detection task. Furthermore, we show the importance and impact of codebook sizing within a Vector Quantized Variational Autoencoder, on the ability of the transformer network to fulfill the task of anomaly detection. Using 294 whole-body PET/CT samples containing various cancer types, we show that our anomaly detection method is robust and capable of achieving accurate cancer localization results even in cases where normal training data is unavailable. In addition, we show the efficacy of this approach on out-of-sample data showcasing the generalizability of this approach even with limited training data. Lastly, we propose to combine model uncertainty with a new kernel density estimation approach, and show that it provides clinically and statistically significant improvements in accuracy and robustness, when compared to the classic residual-based anomaly maps. Overall, a superior performance is demonstrated against leading state-of-the-art alternatives, drawing attention to the potential of these approaches.
\end{abstract}



\section{Introduction}
\label{intro}

\begin{figure}[h]
\centering
\includegraphics[width=8cm]{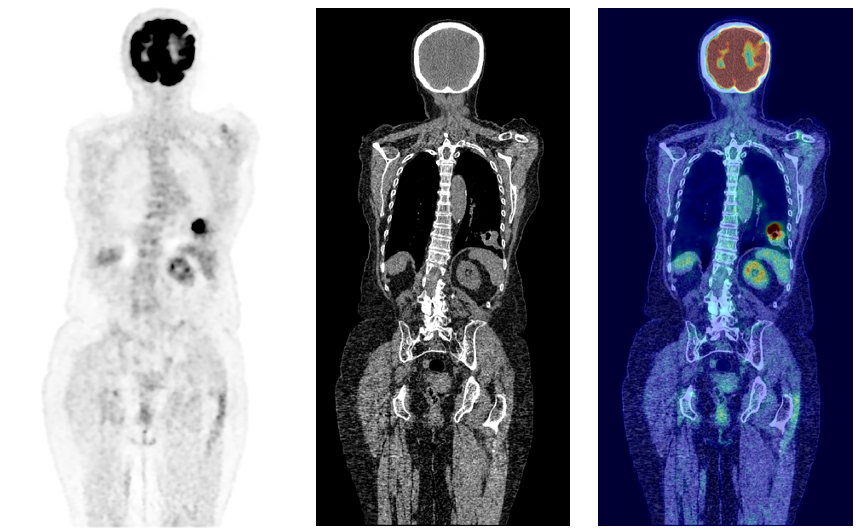}
\caption{\textsuperscript{18}F-FDG PET/CT Imaging. PET (left), CT (centre), combined CT-PET (right). High intensity area in lower chest seen in PET and CT-PET scan shows cancer. Note the non-cancer related high PET signal in the brain and kidney regions.} \label{ct_pet_images}
\end{figure}

Cancer is a disease affecting approximately one in two people over their lifetime \citep{Ahmad2015}. In 2020 alone, over 19 million new cases were reported worldwide, a figure expected to rise by over 50\% by 2040 \citep{Sung2021}. Although preventative measures can be taken through dietary and lifestyle changes, often the first real line of defense is through early diagnosis of which medical imaging plays a key role. Amongst imaging modalities, \textsuperscript{18}F-fluorodeoxyglucose Positron Emission Tomography (\textsuperscript{18}F-FDG PET) has one of the highest detection rates for cancer \citep{Liu2017, Endo2006}. Through enabling the visualization of the glycolytic pathway, the efficacy of \textsuperscript{18}F-FDG PET is related to the high metabolic rates of cancer cells \citep{Almuhaideb2011}. As such PET may enhance cancer staging, treatment planning, and the evaluation of patient responses to treatment \citep{Kim2015}. PET is almost always coupled with CT, and more recently hybrid PET/MRI scanners have been introduced into clinical practice. This allows anatomical localization of \textsuperscript{18}F-FDG uptake as well as attenuation correction of the PET signal. From an unsupervised anomaly detection approach, using CT as an anatomical point of reference by combining modalities can further enhance PET interpretability, exemplified in Fig \ref{ct_pet_images}.

In clinical practice PET images are usually read in a qualitative manner, across a range of clinically relevant tasks from staging, treatment planning, and surgical or therapy intervention planning. Sensitivities can range as much as 35\% for PET imaging depending on the nuclear medicine/PET physician and cancer type \citep{Newman2021}. This can be an issue in the case of metastatic cancer where small lesions can be overlooked \citep{Perani2014}. Considering these shortfalls, there is significant motivation for developing accurate automated detection methods, a major topic of interest in medical imaging research.

\subsection{Related Work}
\label{related_work}
Quantitative imaging analysis is an approach aimed to tackle this problem through segregating normal and pathological findings by finding optimal thresholds. This approach can be done via regional or voxel-wise analysis. For regional analysis, uptake is compared to the same regional uptake found in a healthy control population. This analysis however requires extensive prior knowledge of the subject and control population in order to select the most valid atlases and relevant discriminant regions greatly limiting this approach’s efficacy \citep{Signorini1999}. In classic voxel-wise analysis, a PET image is often registered to a normal standardised group space to compare voxel-wise differences in uptake. Such approaches have been implemented in Neurostat and NeuroGam \citep{Drzezga2005, Renard2013}. Similar implementations have been carried out to generate Z-score maps to demonstrate the degree of abnormality between the healthy model and individual subjects. This approach implemented in \cite{Burgos2021} developed patient-specific models through using a healthy control atlas database. However, these tools have primarily been developed for brain imaging with limited further use outside this area, in addition the approach has sub-optimal assumptions about the statistical occurrence of abnormalities, i.e. noise models and parametric distributions.
\\

Breakthroughs in recent years have primarily showcased the efficacy of deep learning models for anomaly detection in medical data. Given the limitations of previous approaches on whole-body data, the use of deep learning shows promise in the task of anomaly detection.
\\

Unsupervised methods have become an increasingly prominent field in recent years for automatic anomaly detection by eliminating the necessity of acquiring accurately labelled data and showing a strong ability to generalise to unseen anomalies \citep{Chen2020, Baur2020}. These methods mainly rely on creating generative models trained solely on non-anomalous data. Then during inference, anomalies are defined as deviations from the defined model of normality as learnt during training. However, their efficacy is often limited by the requirement of uncontaminated (e.g., non-anomalous) data with minimal anomalies present during training. 
\\

One particular model used for unsupervised anomaly detection is the family of generative adversarial networks (GANs). GANs are able to generate data without explicitly modelling the probability density function of the underlying data \citep{Yi2019}. The overall architecture consists of two sub-networks, namely a discriminator and generator. During training, the generator is tasked with generating samples from a given input (usually gaussian noise). The objective is to generate samples that the discriminator cannot differentiate from real. The use of GANs has many applications in medical imaging, including reconstruction \citep{Quan2018}, denoising \citep{Armanious2018, Kang2018} and cross modality translation \citep{Armanious2018, Bi2017, BenCohen2017,Armanious2019}. GANs also show promise in the field of anomaly detection, as demonstrated in \citep{Schlegl2017, Alex2017, Sun2018}. The approach taken by \cite{Sun2018} for brain MRI anomaly detection used CycleGan, trained to generate healthy-looking scans from anomalous ones. The approach however was shown to be imperfect brought down to textual differences and ununiform intensities during the reconstruction of abnormalities in addition to the instabilities inherent in training GANs models, making them more prone to model collapse \citep{Kodali2017}. As such implementing a GAN architecture poses a greater challenge for the task of 3D image synthesis.
\\

The most prevalent competing family of networks for unsupervised anomaly detection is the Autoencoder (AE). AEs are models that are made of two main architectures, an encoder and decoder. The encoder maps the input image into a lower dimension manifold $z$, where the decoder will then reconstruct the original image from said manifold. The driving characteristic of AEs that make them suitable for anomaly detection is the low dimensional manifold bottleneck; for an AE sufficiently constrained and trained on healthy data only, the model will struggle to generalise when faced with unseen anomalies and will have large reconstruction errors associated at these locations \citep{Baur2020}. Such approaches have been used in various applications, such as brain tumour detection in MRI and CT \citep{Atlason2018, Baur2019, SatoD2018}. However, a common problem with conventional AEs is the lack of regularisation of the latent space, reducing its efficacy for anomaly detection \citep{ChenX2018}. Furthermore, AEs are simply a learnt function whose purpose is to compress data and minimise reconstruction loss. As such, an AE can easily overfit data unless explicitly regularised and can render irregular reconstructions on unseen data.
Improving on these AE limitations, spatial VAEs have been proposed \cite{Baur2020}. Here, the healthy data manifold is obtained by constraining the latent space to conform to a given distribution. A comparative study has shown that VAEs demonstrated improved performance as an alternative to AEs \citep{Baur2020}. The same study also revealed an improvement against adversarial-based strategies like f-AnoGAN \citep{Schlegl2017}. Even with such improvements however, a particular weakness, as explored in \cite{Makhzani2015}, is that the KL-divergence prompts the posterior to incorporate the mode of the prior, not necessarily the entire distribution. This can result in an oversimplified prior brought about by the KL-divergence term resulting in over-regularization. Furthermore, the objective of the VAE can result in trivial solutions that decouple the input from the model’s latent space resulting in posterior collapse \citep{ChenX2016}. This approach is further limited by low fidelity reconstructions and unwanted reconstructions of unseen pathologies suggesting a shortfall in the model itself.
\\

To overcome some of these issues, an approach for unsupervised anomaly detection was presented utilising autoregressive models coupled with a superior autoencoder model, namely the vector-quantised variational autoencoder (VQ-VAE) \citep{Oord2017, Marimont2020}. Transformers, currently state-of-the-art networks in the language modelling domain \citep{Vaswani2017, Radford2018}, use attention mechanisms to learn contextual dependencies regardless of location, allowing the model to learn long-distance relationships to capture the sequential nature of input sequences. This general approach can be generalised to any sequential data, and many breakthroughs have seen the application of transformers in computer vision tasks from image classification to image and video synthesis \citep{Chen2020, Child2019, Yan2021}. Although having showcased state-of-the-art performance in unsupervised anomaly detection tasks for medical imaging data \citep{Pinaya2021}, these methods still rely heavily on purely normal data for model training. To date little research has been carried out using unlabelled (image-wise or pixel-wise) training data that contain anomalies. The work in \cite{Zhang2022} and \cite{Zuluaga2011} proposes methods that make use of anomalous training data, but even so for this method to work, an initial portion of labelled normal training data is required to successfully make use of the unlabelled training portion. To the best of our knowledge, no prior research exists using unsupervised methods to accurately localise anomalies while using training data containing such anomalies with no prior knowledge over any samples whether they they contain anomalies or not. This task itself is of importance given the nature of whole-body PET. It is often difficult or unethical to obtain healthy datasets of certain medical imaging modalities as some images are only acquired with prior suspicion of disease.

\subsection{Contributions}
\label{contributions}

To address these problems and shortfalls of existing state-of-the-art techniques, we propose a method for unsupervised anomaly detection and segmentation using multi-model imaging via transformers with cross attention. Leveraging off previous work combining the use of VQ-VAE models with transformers we propose, deploy and evaluate the following contributions in our work:

\begin{itemize}
  \item We show the added benefit of the use of multi-modal imaging for unsupervised anomaly detection, achieved through the use of transformers with cross attention
  \item We highlight the importance of optimal choices in the VQ-VAE codebook architecture, beyond accurate reconstruction performance, and its effect further downstream on the transformer’s performance for anomaly detection
  \item We introduce an improved alternative to commonly used residual-based anomaly maps via a kernel density estimation approach
  \item We supplement this kernel density estimation approach with an extensive study of kernel choices and a selection of regularisation parameters.
  \item We carry out our training on data where healthy samples are unattainable and still show high detection rates during testing where other models fail
\end{itemize}

This work is an extended version of a conference workshop paper presented at \\
DGM4MICCAI \citep{Patel2022}. The extensions to the
conference paper involve a more substantial literature review in addition to a larger dataset that has yielded a greater number of training samples and a higher number of samples for testing on unseen cases. Additionally a comprehensive ablation
study showcasing the importance of codebook sizing in the Vector Quantized-Variational Autoencoder is explored
that can highlight its affect on the performance of the transformer model for anomaly detection. A deeper exploration into the kernel density estimation approach is also carried out showcasing the difference in performance of varying kernel choices in addition to varying regularisation parameters for this methodology. Finally, as further validation of our methods, we carry out an additional set of testing on a fully out-of-sample testing dataset with varying cancer cases from a different source to that of the original training and testing data.

\section{Proposed Method}
\label{method}

The principal components behind the proposed whole-body anomaly detection model relies on using transformer models and auto-encoders to model 3D whole-body \textsuperscript{18}F-FDG PET scans. Although all training data contain anomalies, the spatial distribution of anomalies across samples will result in such anomalies being unlikely, thus appearing at the likelihood tail-end of the learnt distribution. In order to use transformer models, images need to be expressed as a sequence of values, ideally categorical. As it is not computationally feasible to do this using voxel values, a compact quantized (discrete) latent space is used as input for the transformer via a VQ-GAN model as named and proposed in \cite{Esser2020} (a VQ-VAE \cite{Oord2017} with an adversarial component).

\subsection{VQ-GAN}
\label{vq-gan}

The original VQ-VAE model \citep{Oord2017} is an autoencoder that learns discrete latent representations of images. The model comprises of three principal modules: the encoder that maps a given sample $X \in \mathbb{R}^{H\times W\times D}$ onto a latent embedding space $e \in \mathbb{R}^{h\times w\times d\times n_{z}}$ where $n_z$ is the dimension of each latent vector $e_i$. After the encoder network projects the image $X$ to its latent representation $z_e(X)$, the discrete latent variables $z$ are generated by a nearest neighbour look-up to the shared embedding space $e$ to generate $e_m, m \in 1,...M$ where $M$ is the vocabulary size. For simplicity we can refer to a single random variable as $z$ to represent a single discrete latent variable, given the input this can represent a 1D, 2D or 3D latent feature space. In this case given we are dealing with 3D medical data $z$ corresponds to a 3D feature space. During training the codebook is learnt jointly with model parameters. The posterior distribution can then be given as a categorical one defined as:

\begin{equation}\label{eq:quantization}
q(z=m|X) = \left\{\begin{matrix}
1\ for \: m = \text{argmin}_{j}\left \| z_e(X)-e_{j} \right \|_{2}
\\ 
0\ otherwise
\end{matrix}\right.
\end{equation}

Here $z_e(X)$ is the output of the encoder giving us our encoded feature vectors that are matched to $e_j$, a learnt codebook vector in the shared embedding space. The discrete latent space representation is thus a sequence of indexes for each code from the codebook. The final portion of the network is the decoder, which reconstructs the original observation from the quantized latent space. The total loss for the VQ-VAE is then given as:
\begin{equation}
    \resizebox{0.9\columnwidth}{!}{$
L_{VQVAE} = \left \|(X-\hat{X})  \right \|^2_{2} + \left \| \left | STFT(X) \right | - \left | STFT(\hat{X}) \right |\right \|^2_{2}  + \left \| sg[z_e(X)]-e \right \|_{2}^{2}  + \beta \left \| z_e(X) - sg[e]]\right \|_{2}^{2} $}%
\label{eq:jukebox}
\end{equation}

\noindent where $\hat{X}$ is the output from the decoder and $sg$ is a stop gradient operator to stop gradients from flowing back into their argument. The loss function for the VQ-VAE makes use of a spectral loss \citep{Dhariwal2020} that is, it includes a component based on the magnitude of the Fourier transform of the original and reconstructed image. From equation \ref{eq:jukebox} the first term is the pixel loss, the second term is the spectral loss between the original and reconstruction where SFTF stands for the short time Fourier transform. The third term is the commitment cost used to ensure the encoder commits to the codebook. The final term is to move the codebook embedding vectors towards the output from the encoder. For this term, we replace this and use the exponential moving average updates for the codebook \citep{Oord2017}. During training, a $\beta$ of 0.25 was used.
\\

As autoencoders often have limited fidelity reconstructions \citep{Dumoulin2016}, an adversarial loss is added to the VQ-VAE network to form a VQ-GAN, as proposed and named in \cite{Esser2020}. When implementing the VQ-GAN network, however, due to instabilities associated with adversarial networks, the loss function is further expanded to include a perceptual loss \citep{Takaki2018} that helps preserve spatial consistency by using the LPIPS library \citep{Zhang2018}. The perceptual loss applied via the LPIPS library makes use of 2D images and as such has to be applied over slices of the original and reconstructed sample. To improve efficiency during training the loss is randomly applied to 50\% of slices across each plane. The adversarial loss used is of the following form:

\begin{equation}
\mathcal{L}_{adv}=\mathcal{L}_{LSGAN}(Dis)+\mathcal{L}_{LSGAN}(E,Quant,D)
\label{eq:adversarial}
\end{equation}

Where $Dis$ is the discriminator and $E,Quant,D$ is the VQ-VAE Encoder, Quantizer and Decoder Respectively. This is based on the Patch-GAN model \citep{Isola2017} as per \citep{Esser2020} and paired with the LS-GAN loss \citep{Mao2017} providing a more stable and reproducible behaviour, denoted as: 

\begin{align*}
    \min _{Dis} {L}_{\text{LSGAN}}(Dis)=&\frac{1}{2} \mathbb{E}_{\mathbf{x} \sim p_{\mathrm{data}}(\mathbf{x})}\left[(Dis(X)-1)^{2}\right]\\
    &+\frac{1}{2} \mathbb{E}_{\mathbf{x} \sim p_{\mathbf{data}}(\mathbf{x})}\left[(Dis(D(Quant(E(X))))^{2}\right]
    \\
    \min _{Enc,Quant,Dec} {L}_{\text{LSGAN}}(E,Quant,D)=&\frac{1}{2} \mathbb{E}_{\mathbf{x} \sim p_{\mathrm{data}}(\mathbf{z})}\left[(Dis(D(Quant(E(X))))-1)^{2}\right]
    \label{eq:ls_gan_dis}
\end{align*}

\subsection{Transformer}
\label{transformer}

Once a VQ-GAN model is trained, we now are required to train a generative model on the discrete latent representation. For this we use a transformer. Transformer models rely on attention mechanisms to capture the relationship between inputs regardless of the distance or positioning relative to each other. The self-attention mechanism is best described as a mapping of intermediate representations of three position-wise linear layers onto three representations denoted by the Value (V), key (K) and query (Q) \citep{Vaswani2017}. With $d_k$ denoting the dimension of the key vectors, the attention mechanism is calculated as:

\begin{equation}\label{eq:Attn}
Attn(Q,K,V) = softmax\left ( \frac{QK^{T}}{\sqrt{d_k}} \right )V
\end{equation}

The multi-head attention aspect of this transformer network is then several attention layers run in parallel with their outputs concatenated and fed through a linear layer.
This process, however, relies on the inner product between elements and, as such, network sizing scales quadratically with sequence length. Given this limitation, achieving full attention with large medical data, even after the VQ-GAN encoding, comes at too high a computational cost. To circumvent this issue, many efficient transformer approximations have been proposed \citep{Tay2020, Choromanski2020}. In this study, a Performer model is used; the Performer uses the FAVOR+ algorithm \citep{Choromanski2020}, which proposes a linear generalized attention that offers a scalable estimate of the attention mechanism. Using such a model, we can apply transformer-like models to much longer sequence lengths associated with whole-body data.
In order to learn from the discrete latent representations, we require the discretised latent space $z_q$ to take the form of a 1D sequence $s$ using some arbitrary ordering. The transformer is then used to model $s$ by minimizing the conditional distribution $p(s_{i})= p(s_{i}\mid s_{< i})$ where $i$ is the $i^{th}$ element of $s$.

\subsection{Anomaly Detection}
\label{AD}
To perform the baseline anomaly detection model on unseen data, as proposed by \cite{Pinaya2021}, first, we obtain the discrete latent representation of a test image using the VQ-GAN model. Next, the latent representation $z_q$ is reshaped using a 3D raster scan into a 1D sequence $s$ where the trained Performer model is used to obtain likelihoods for each latent variable. At each position in the sequence the trained transformer will give the learnt likelihood of each possible token appearing at every point in the sequence. In doing so we can highlight low likelihood (or anomalous tokens) as $p(s_{i})= p(s_{i}\mid s_{< i}) < t$, (where $t$ is a threshold determined empirically using a validation dataset; t = 0.025 was found to be optimal). This generates a binary resampling mask that indicates which tokens in the latent sequence are anomalous i.e. below the threshold. Using the resampling mask, the anomalous latent variables are removed and replaced in the sequence with non-anomalous tokens by resampling from the transformer. This approach replaces anomalous latent variables with those that are more likely to belong to a healthy distribution, as such "healing" the considered anomalous latent space. Using the non-anomalous latent space, the VQ-GAN model reconstructs the original image $X$ as a non-anomalous reconstruction $C_r$. Finally, a voxel-wise residual map can be calculated as $X – X_r$ with final segmentations calculated by thresholding the residual values. As areas of interest in PET occur as elevated uptake, residual maps are filtered to only highlight positive residuals.

\subsection{CT Conditioning}
There are often times when more information can be useful for inference. This can be in the imaging domain through multiple resolutions \citep{Chen2021} or multiple modalities/spectrums \citep{Mohla2020}. It is for these tasks where cross-attention can prove beneficial. From a clinical point of view, whole-body PET scans are acquired in conjunction with CT, or less frequently MRI data for attenuation correction purposes in addition to providing an anatomical reference. Additionally, it can be observed that areas of high uptake are not always associated with pathological findings i.e. high uptake may reflect physiological uptake, e.g. within the brain and heart. Additionally areas where radiotracer may collect like the kidney and bladder can also show high uptake patterns. Acknowledging these areas of high physiological uptake by recognition of the organ location with respect to a whole 3D scan visible may seem obvious to the human eye, however this may not be the case using the transformer approach. For this work images are encoded to a discrete latent space and then rasterized into a 1D sequence. During training and inference the model works in an autoregressive manor i.e. only prior tokens in the sequence can be viewed. As such when looking at a specific token (and only the prior tokens in the sequence), it may be hard for the model to determine the exact anatomical point within the whole body that that specific token represents, without further context relating to the whole body. As such the anatomical reference provided from CT data is beneficial. This leads to one of the main contributions of the work, namely anomaly detection incorporating CT data. This process works by generating a separate VQ-GAN model to reconstruct the PET-registered CT data. Then, CT and PET data are encoded and ordered into a 1D sequence using the same rasterization process, such that CT and PET latent tokens are spatially aligned. The transformer network is then adapted to include cross-attention layers \citep{Gheini2021} that feed in the embedded CT sequence after each self-attention layer. At each point in the PET sequence, the network has a full view of the CT data helping as a structural reference. In doing so, the problem of determining the codebook index at a given position $i$ becomes $p(s_{i})= p(s_{i}\mid s_{< i},c)$, where $c$ is the CT latent sequence. 

\begin{figure*}[h]
\centering
\includegraphics[width=14cm]{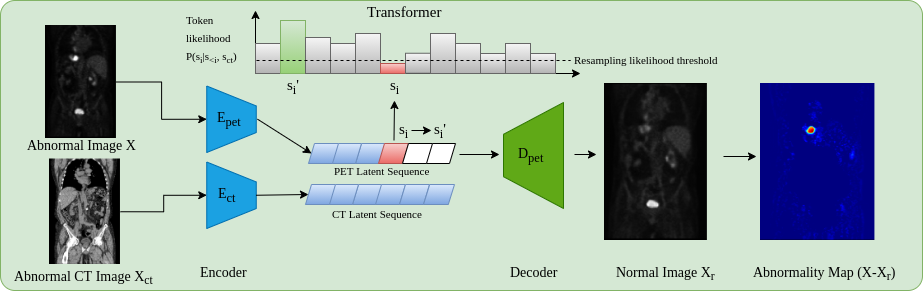}
\caption{Anomaly Detection Pipeline – PET image $X$ is encoded along with CT image $X_{ct}$. Tokens form the encoded PET image are then sampled from the transformer by obtaining their likelihood with respect to prior tokens in the sequence and all CT tokens. Tokens below a given threshold are resampled from a multinomial distribution, derived from likelihood outputs from the transformer for all tokens at a given position in the sequence. This yields a non-anomalous latent space which is decoded to give $X_r$.} \label{ad_pipeline}
\end{figure*}

To add cross attention to the transformer architecture, we add a cross attention layer after each self-attention layer in the transformer architecture. Still using the same attention mechanism, the cross attention calculation is then given as:

\begin{equation}\label{eq:xAttn}
Attn(Q_s,K_c,V_c) = softmax\left ( \frac{Q_sK_c^{T}}{\sqrt{d_k}} \right )V_c
\end{equation}

Where $Q_s$ is the output from the prior self-attention layer, and $K_c$ and $V_c$ are the Key and Query vectors derived from the embedded conditioning CT sequence. The architecture of an entire transformer layer with self-attention and cross-attention can be visualised as in Fig. \ref{cross_attention}:

\begin{figure}[h]
\centering
\includegraphics[width=8cm]{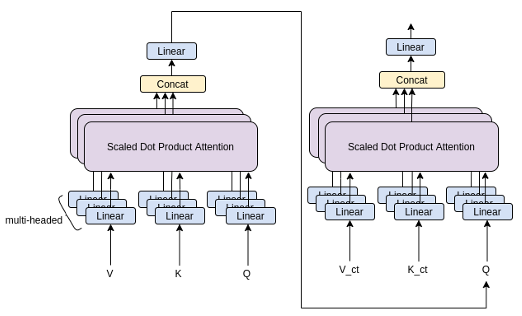}
\caption{Network architecture showcasing layers in transformer with multi-headed self-attention (left) and corresponding multi-headed cross-attention (right) run in series} \label{cross_attention}
\end{figure}

This approach for anomaly detection, as visualised in Fig. \ref{ad_pipeline}, adds robustness to the anomaly detection framework by providing meaningful context in areas of greater variability in uptake that can be explained by the anatomical information within CT.

\subsection{Kernel Density Estimation}

A drawback of the baseline anomaly detection method described in section 2.3 is that the residual image uses an explicit image-wide threshold to generate a segmentation map. The resulting segmentation can often be noisy due to discrepancies between the reconstructed image and the original, for example, between borders of high intensity. Additionally, anomalies can occur at different intensities, meaning a blanket threshold is inappropriate. A possible solution is to implement Z-score anomaly maps as used in similar anomaly detection work \citep{Burgos2021}. For this work, this can be achieved by introducing stochasticity within the model. As the distribution of uptake patterns is often multi-modal, due to its relationship to base metabolic rate and procedure-related variations (eg injected tracer amount and time since injection), the optimality of the Z-score’s Gaussian-error assumption should be questioned and relaxed. Empirical evidence obtained by exploring the data and by sampling from the transformer itself highlights that the error is indeed non-Gaussian even in non-anomalous regions. For example, in the heart, bi-modal (even multi-modal) error distributions are observed. To remedy this, we propose to use a non-parametric approach using kernel density estimation (KDE) \citep{Parzen1962}. To do this, we sample from the model by introducing a dropout layer in the VQ-GAN decoder. Additionally, we achieve further stochasticity by replacing unlikely tokens with ones drawn from a multinomial distribution, derived from the likelihoods output from the transformer for each token at a given position in the sequence. During inference, by sampling multiple times, we generate multiple normal latent representations for a single image, which are then decoded multiple times with dropout to generate multiple non-anomalous reconstructions of a sample, at which point a KDE is fit independently at each voxel position to generate an estimate of the probability density function $f$ for the intensities at a specific point across reconstructions. Letting $(x_1, . . . , x_n)$ be the intensity for a voxel position across reconstructions, we can generate an estimation for the shape of the density function $f$ for voxel $x$ as:

\begin{equation}\label{eq:kde}
\hat{f_{h}}(x)=\frac{1}{nh}\sum_{i=1}^{n}K\left ( \frac{x-x_{i}}{h} \right )
\end{equation}

Here, $K$ is a given kernel shape and $h$ is a smoothing bandwidth calculated via the Silverman method \citep{Silverman2018} as:

\begin{equation}\label{eq:bandwidth}
h=\left ( \frac{4\hat{\sigma}^{5}}{3n} \right )^{1/5} + \epsilon
\end{equation}

\noindent with $\hat{\sigma}$ representing the standard deviation at the given voxel position across $n$ reconstructions and $\epsilon$ is a scalar regularisation parameter determined empirically on the validation dataset used to address areas of low variance across reconstructions. We can then score voxels from that estimated density function at the intensity of the real image, at the voxel level, to generate a log-likelihood for that intensity, generating the anomaly map.
The KDE approach can be further explored via varying kernels and regularisation parameters. As part of this study, we evaluate the use of a series of different Kernels to calculate the estimation of the probability density function $f$ in addition to varying regularisation values for $\epsilon$.

\subsection{Clinically Consistent PET Segmentations}

For whole-body PET, due to the large system Point Spread Function (PSF) of observed uptake, the contours of an anomaly can be hard to define. Given this there are a number of methods used clinically to define the boundaries of anomalies. This can range from absolute thresholding over the entire image, to per anomaly thresholding via percentages of the maximum uptake of individual anomalies, to more involved and computationally expensive algorithms \citep{Berthon2017,Hatt2017}. In general there is no universal standard for defining the anomaly boarders and as such we use the method most recognised as the clinical standard in the UK as advised by our clinical experts. The method used makes use of percentage thresholding for individual anomalies. This defines boundaries of an anomaly as connecting voxels with intensities above 40\% of the maximum intensity of a specific anomaly and has shown to generate optimal performance with limited computation \citep{Berthon2017}. To conform to this standard, we apply a final post-processing step of growing all initial segmentations to satisfy this criteria.

\section{Data}
\label{data}

For the training, validation and testing of the methods described above, a combination of two datasets was used to overcome the limitations of limited data for training purposes and the lack of validity for the proposed methods during testing in the case of limited testing samples. Furthermore, to improve the efficacy of the approach to unseen out-of-distribution data, the approach will likely be supplemented with training from data that have undergone different acquisition protocols and varying voxel dimensionality. 
All data were rigidly registered to a group-wise space with a field of view from the neck down to the upper thigh region. The initial step of the group-wise registration was to register all samples to a given sample in the private dataset meaning all datasets were converted to the same voxel spacing, after which, all CT images were registered to their paired PET image using rigid transformations. All processed images had a final dimension of $168\times208\times216$. CT data were further preprocessed to remove the bed from the images and have voxel intensities clipped to showcase the soft-tissue window only.

\subsection{Private dataset}
The private dataset used consists of 83 co-registered \textsuperscript{18}F-FDG PET/CT images acquired using a GE Medical Systems scanner. The original PET images had voxel dimensions of $2.7\times2.7\times3.3mm$, whilst the CT images had dimensions of $1.4\times1.4\times3.3mm$. The dataset comprised a variety of subjects showcasing various primary cancers in various locations across the body and metastases located in numerous further locations. Out of the 83 samples, 60 were used for training whilst the remaining 23 were split, with 12 used for validation and 11 used for testing. From the training data, a total of 49 cases showcased some form of cancer in the scan.

\subsection{NSCLC Radiogenomics Dataset}
The NSCLC Radiogenomics dataset comprises 211 co-registered \textsuperscript{18}F-FDG PET/CT samples presenting Non-small cell lung cancer cases \citep{Bakr2017, Bakr2018, Gevaert2012, Clark2013}. The acquisition protocol ranges by candidate using both GE Medical Systems and Siemens Scanners. The original PET images had voxel dimensions of $3.6\times3.6\times3.3mm$, while the CT images had dimensions of $1.4\times1.4\times3.3mm$. Out of the 211 samples, 160 were used for training, with  26 used for validation and 25 used for testing. From this training data, all cases contained some form of lung cancer along with potential metastases.
\\

From combining the private dataset and NSCLC Radiogenomics dataset, during training, a total of 220 samples were used, of which 209 samples had some form of anomaly present in the scan. Furthermore 38 samples were used for validation to tune hyperparameters and run ablation studies on model parameters like codebook sizing and Kernel types. The remaining 36 samples were left as a hold-out set for testing on the final models chosen after all ablation studies were carried out. 

\subsection{AutoPET Dataset}
For a comparison of our methods and baselines when trained on fully normal data we leverage the autoPET dataset -  a 3D whole-body Positron Emission Tomography (PET) dataset ~\citep{AutoPET2022,TCIA2013}. Generally speaking in PET imaging, scans void of any forms of anomalies are hard to come by, as often scans are taken with a strong prior suspicion of a pathology. There are some cases where this may not be true, that includes scans following treatment, which is where the normal samples for this dataset are obtained. This dataset consists of 1014 PET scans with 430 non-anomalous scans, with the remaining containing some form of lung cancer, lymphoma, or melanoma. From this dataset we generate a separate training dataset with normal cases only. We use all 430 healthy scans to form the training data. For this work the same validation set consisting of the private data and NSCLC radiogenomics data is used to tune the model and anomaly detection hyperparameters, as testing is carried out on their testing set as well for the most fair comparison. The original PET images have voxel dimensions of $2.036\times2.036\times3mm$. All scans are aligned to the same space as the training data via affine transformations, i.e. a field of view from the neck region to upper thigh region is used out of the whole body images.

\subsection{Clinical Proteomic Tumor Analysis Consortium - CPTAC}
The CPTAC dataset comprises a combination of several individual CPTAC studies that cover numerous cancers from lung, ovarian, pancreas and skin cancer \citep{CPTAC2018, Clark2013}. A total of 14 co-registered whole-body \textsuperscript{18}F-FDG PET/CT images are used to demonstrate the ability of the proposed methods on fully out-of-sample data that is not used during training or validation. The data is similarly acquired using a GE Medical Systems Scanner. The original PET images had voxel dimensions of $3.6\times3.6\times3.3mm$, whilst the CT images had dimensions of $1.4\times1.4\times3.3mm$. All scans are registered to the same space as the training data.

\section{Experiments and Results}

\subsection{VQ-GAN Training and architecture details}
The training details and architecture for the VQ-GAN (besides codebook sizing) remains the same through all experiments run in this study. The architecture used for the VQ-GAN model uses an encoder consisting of three strided convolutional layers with stride 2 and kernel size 4. Each convolutional layer is then followed by a ReLU activation and 3 residual blocks (consisting of a 3x3x3 conv, ReLU, 1x1x1 conv, ReLU). The decoder similarly has 3 residual blocks, each followed by a transposed convolutional layer with stride 2 and kernel size 4. Finally, before the last transposed convolutional layer, a Dropout layer with a probability of 0.05 is added.  Further hyperparameters include the use of $\beta$ equal to 0.25, as stated in equation \ref{eq:jukebox}. This value is taken from the original implementation of the VQ-VAE as stated in \cite{Oord2017}.
Through the ablation study exploring the effect of ranging codebook sizes, several codebook dimensions were explored, consisting of atomic elements from 64-2048 with lengths ranging from 32-256. However, changes in the codebook parameters had no change to the encoder and decoder architecture. 
To train the VQ-GANs, we used an ADAM optimiser \citep{Kingma2014} with a learning rate of 1e-4 and an exponential learning rate decay with a gamma of 0.9999. Additionally, the discriminator network had a learning rate of 5e-4. Training data was augmented using elastic deformations, Gaussian noise, intensity shifts, contrast adjustments and gaussian blur. The model was trained over 1000 epochs with a batch size of 3. Further details on model complexity and the number of model parameters can be seen in table \ref{tab:num_params} in Appendix \ref{model_params}.

\subsection{Transformer Training and architecture details}

The performer in all experiments used corresponds to a decoder transformer architecture with 16 layers, each with 8 heads and an embedding size of 256. 
To train the performer network, we used an ADAM optimiser \citep{Kingma2014} with a learning rate of 1e-3 and an exponential learning rate decay with a gamma of 0.9999. The loss function used for training was cross-entropy, given the discrete nature of the latent sequence codes. 
To obtain the input data for training the transformer network, the trained VQ-GAN model encoded the training data into the discrete latent codes, which were used as inputs for the transformer. To avoid overfitting, the original training dataset was augmented 4 times and then encoded using elastic deformations, Gaussian noise, intensity shifts, contrast adjustments and gaussian blur to increase the number of samples for training. The model was then trained over 120 epochs with a batch size of 1. For the AutoPET dataset the models were trained for 80 epochs with a batch size of 1. Further details on model complexity and the number of model parameters for the Performer model with and without cross attention can be seen in table \ref{tab:num_params} in Appendix \ref{model_params}.

\subsection{Experiment 1: PET-only ablation study}

\begin{figure}[h]
\centering
\includegraphics[width=8cm]{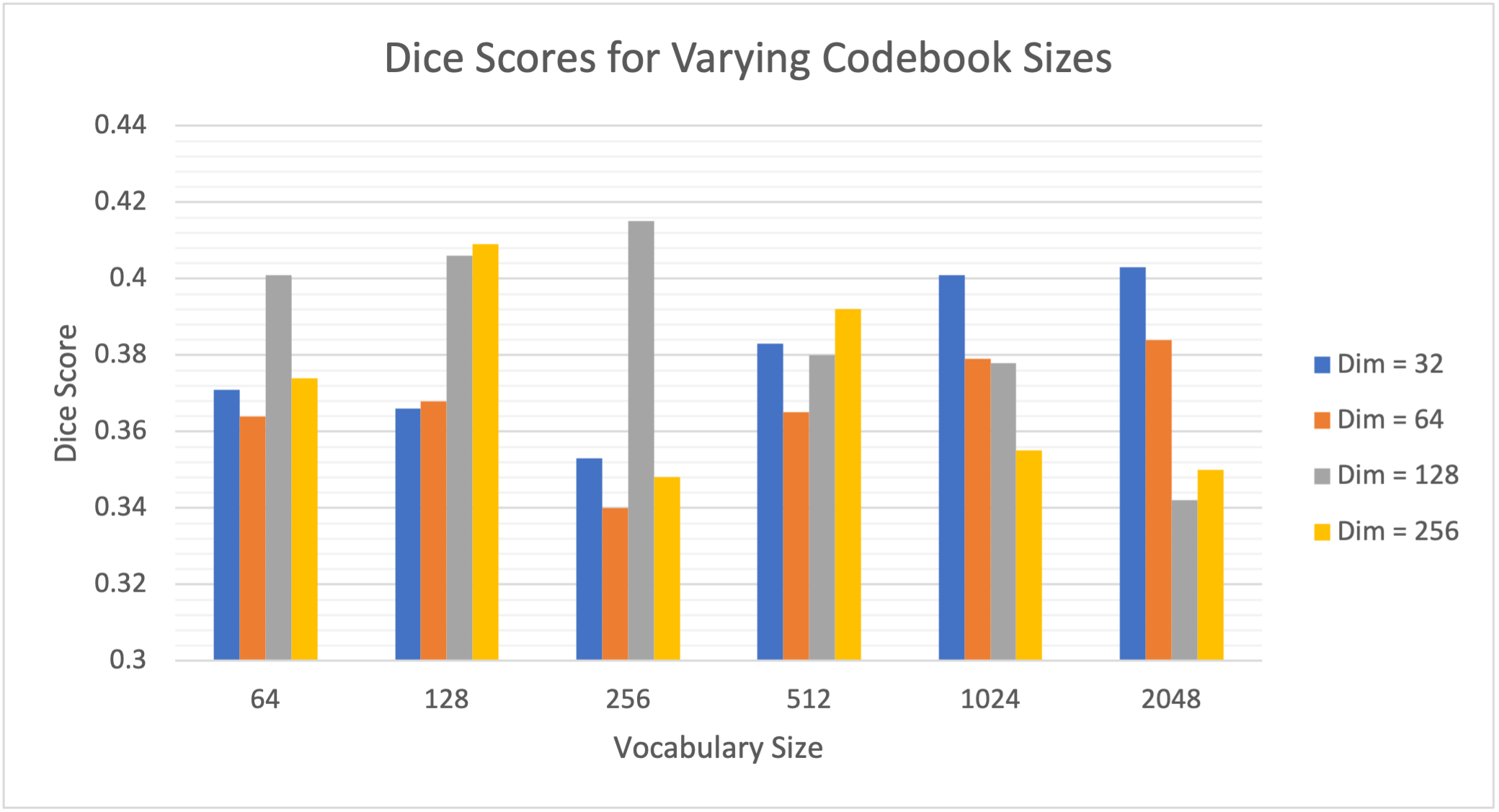}
\caption{Best achievable DICE-score ($\lceil DICE \rceil$) on validation data for anomaly detection with PET data only and varying codebook vocabulary sizes and latent code dimensions} \label{pet_ablation}
\end{figure}

The first study explores the effect of codebook size on the anomaly detection pipeline. To do this, we experiment with codebooks ranging vocabulary sizes from 64-2048 and vector dimensions from 32-256. From this range, 24 different VQ-GAN networks of ranging codebooks were trained, of which a further 24 transformer networks were trained using their respective VQ-GAN model using PET data only. Using the trained networks, anomaly detection was run on the validation dataset where anomalies were located via residual maps between the original and reconstructed images. We measure our models’ performance using the best achievable DICE score, which serves as a theoretical upper-bound to the models’ segmentation performance. This work makes use of the anomalous training dataset. The results can be visualised in Fig. \ref{pet_ablation}

We can see a large range (0.332-0.415) in DICE scores from adjusting the codebook parameters highlighting the importance of choosing an optimal codebook size for anomaly detection. Furthermore, a general trend can be seen from the results showcasing that codebooks with a smaller vocabulary size generally perform better with large dimensions, whereas when the vocabulary size increases, the network generally performs better with smaller latent vector dimensions. From the results, however, a codebook with a vocabulary size of 256 with dimension 128 showed to generate the best results.

\subsection{Experiment 2: CT Conditioning Ablation Study}

Continuing with showcasing the importance of the codebook size for anomaly detection, we further explore ranging codebook sizes with respect to the conditioning sequence from the CT data with results showcased in Table \ref{exp2_results}. For the experiment, we generate 4 different VQ-GAN models to encode CT images with ranging codebook sizes, chosen and conforming to the results seen in Experiment 1, i.e., small vocabulary size with large dimensions and large vocabulary sizes with smaller dimensions. For each of these experiments, we use a PET VQ-GAN network with the optimal codebook size as derived from Experiment 1, i.e., vocabulary size of 256 with dimension 128. As with Experiment 1, we train the models on the anomalous training dataset and run anomaly detection on the validation set to generate DICE scores using residual maps. As in Experiment 1, a vocabulary size of 256 and dimension of 128 was found to produce the best dice score.

\begin{table}[h]
\centering
\caption{Anomaly detection results on whole-body PET validation data for varying codebook sizes. The performance is measured with best achievable DICE-score ($\lceil DICE \rceil$) and AUPRC on the test set.}\label{exp2_results}
\resizebox{6cm}{!}{%
\begin{tabular}{lcc}
\hline
 Vocabulary Size & Dimensions & \bfseries $\lceil DICE \rceil$  \\
  \hline
64              & 256       & 0.424 \\
256             & 128       & 0.458 \\
512             & 64        & 0.432 \\
1024            & 32        & 0.443 \\
\hline
\end{tabular}}
\end{table}

\subsection{Experiment 3: KDE Ablation Study}

When implementing the kernel density estimation approach, the shape of the probability density estimation relies on 2 factors, the shape of the kernel used and the bandwidth. Given this, we explore varying kernels in addition to a range of $\epsilon$ values (the regularisation term as seen in equation \ref{eq:bandwidth}). In total, six  different kernels are used (i.e. gaussian, top-hat, Epanechikov, exponential, linear and cosine), of which their shape can be seen below in Fig. \ref{kernels}. 

\begin{figure}[h]
\centering
\includegraphics[width=8cm]{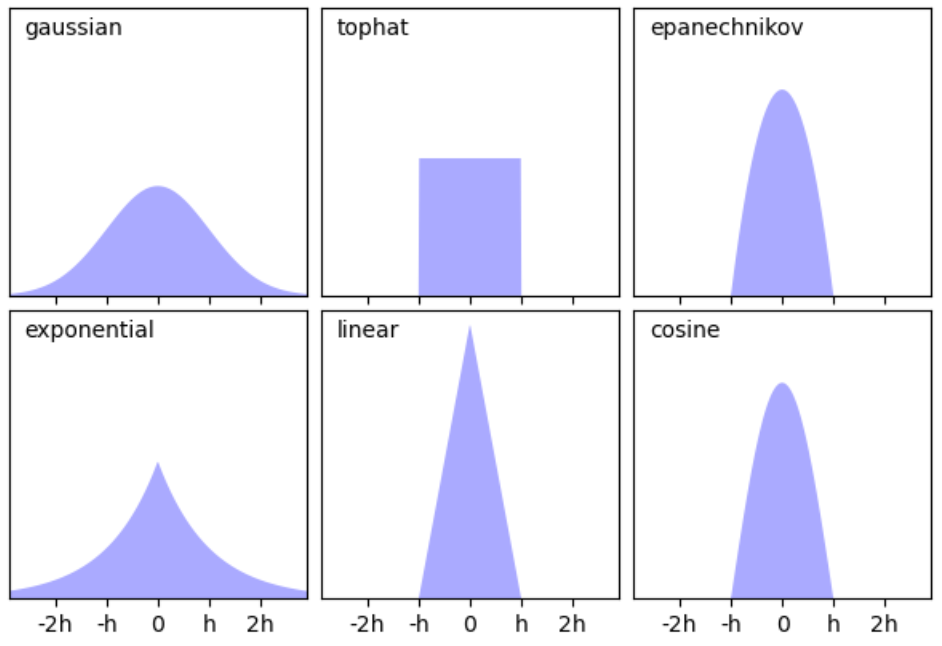}
\caption{Different Kernel shapes with respect to bandwidth (h) explored in KDE ablation study (\cite{scikit_kde})} \label{kernels}
\end{figure}

Furthermore, three $\epsilon$ values of 0.025, 0.05 and 0.1 were experimented with. From experiments 1 and 2, we chose the best vocabulary sizes for the PET VQ-GAN and CT VQ-GAN, which is a vocabulary size of 256 and dimension of 128 for both models. As we use the kernel density estimation approach, not residual maps, we rely on multiple reconstructions achieved through resampling from the transformer 60 times, with each latent sequence decoded with dropout four times each. As with the previous experiments score, we measure our models’ performance using the best achievable DICE score based on the KDE anomaly maps.

\begin{figure}[h]
\centering
\includegraphics[width=8cm]{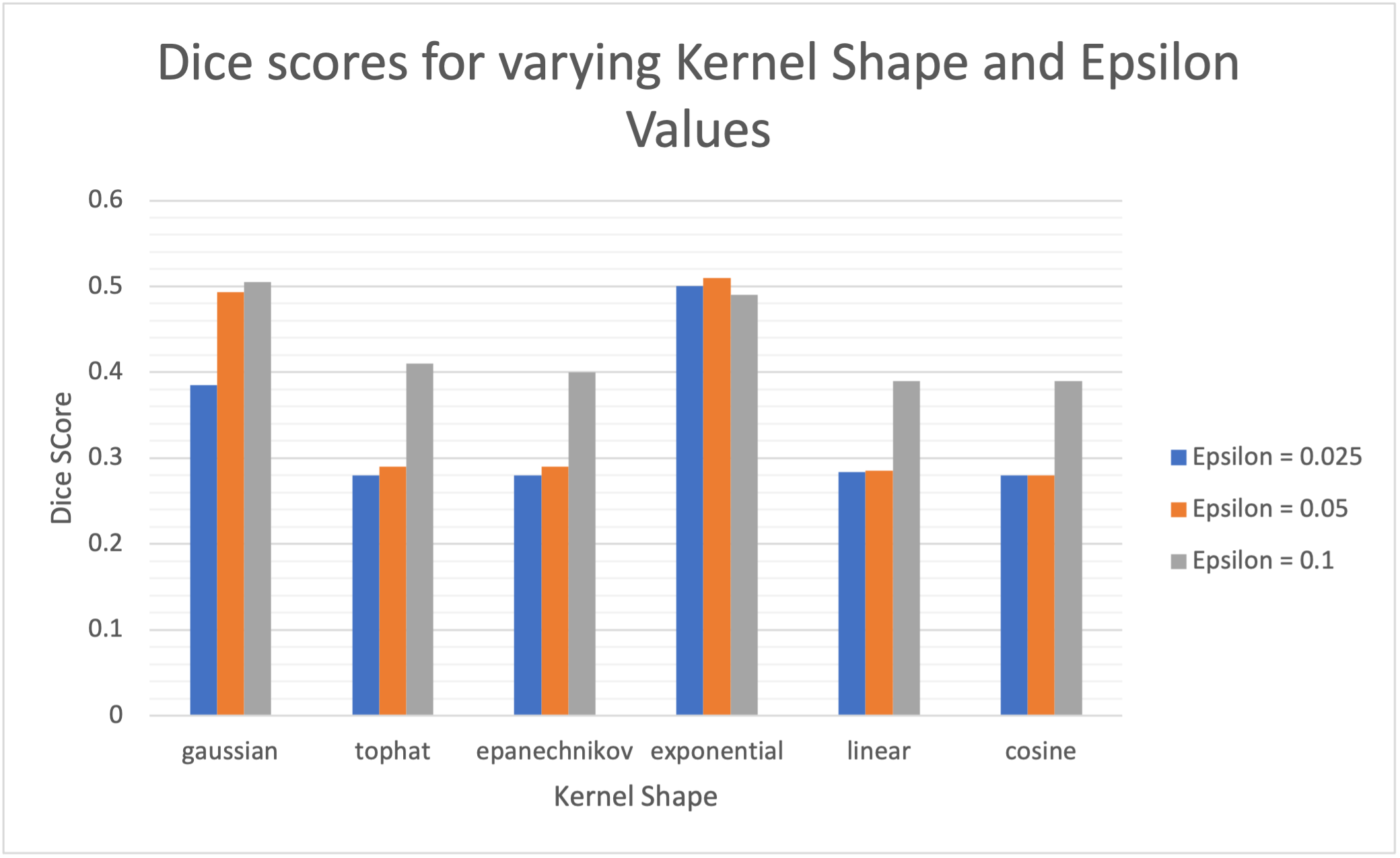}
\caption{Best achievable DICE-score ($\lceil DICE \rceil$) on validation data for Kernel Density Estimation approach with varying Kernel shapes and $\epsilon$ values} \label{kernels_results}
\end{figure}

From the results, there are two clear standout kernels that showcase the best performance, namely the gaussian and exponential kernel. The shape of the kernels implies that kernels with heavier tails seem to be more favorable for the task. Additionally, we generally see higher $\epsilon$ values, and therefore higher bandwidths and longer tails, give better performances, with large reductions in performance associated with the smallest $\epsilon$ values due to under regularisation and noisy KDE maps. Furthermore, we see the exponential kernel produces the best results in addition to the least variance associated with ranging $\epsilon$ values indicating less reliance on an optimal choice of $\epsilon$ and perhaps a more robust approach when facing unseen data.
\\

From the series of experiments optimized on the validation data, a final model is generated using a PET and CT VQ-GAN model with a codebook vocabulary size of 256 and dimension 128. Furthermore, on the kernel density estimation side, final parameters are set to calculate probability density estimates using an exponential kernel with an $\epsilon$ value of 0.05.

\subsection{Testing}
\label{Testing1}
After obtaining the optimal models from each hyperparameter study, we test the proposed models on the hold-out testing data as a final ablation study to showcase the added contribution of each proposed method. We compare our results to that of a Dense, Spatial AE and Dense VAE model \citep{Baur2020} and an SSIM AE \citep{Bergmann2019}. These models are trained on PET only, however we also implement them with 2 channel inputs with PET and CT to showcase the effect of CT conditioning on the baseline models in comparison to the effects of conditioning for our proposed method.  We additionally implement F-AnoGan for a further baseline comparison \citep{Schlegl2017}. Testing is carried out on 36 samples of which the best achievable DICE score is calculated in addition to the area under the precision-recall curve (AUPRC) as a suitable measure for segmentation performance under class imbalance. These results can be seen tabulated in Table \ref{testing_results} and visualised in Fig. \ref{testing_comparisons} qualitatively showcasing the best two state-of-the-art models against our added contributions. The precision recall curves highlighting important model comparisons can also be seen in Appendix \ref{auc_curves}. Furthermore, we carry out paired t-tests to showcase the statistical significance of improvements.

\begin{table*}[h]
\centering
\caption{Anomaly detection results on whole-body PET hold-out testing data. The performance is measured with best achievable DICE-score ($\lceil DICE \rceil$) and AUPRC on the test set.}\label{testing_results}
\resizebox{15cm}{!}{%
\begin{tabular}{lcc|cc}
\hline
\textbf{Method}                                                                                & \multicolumn{2}{c|}{Anomalous Training Data} & \multicolumn{2}{c}{Normal Training Data} \\
\hline
                                                                                               & $\lceil DICE \rceil$              & AUPRC             & $\lceil DICE \rceil$             & AUPRC   
                                                                                               \\
                                                                                               \hline
AE (Dense) \citep{Baur2020}                                                   & 0.333                   & 0.301             & 0.371                  & 0.322            \\
AE (Dense) + CT \citep{Baur2020}                                              & 0.332                   & 0.281             & 0.360                  & 0.319            \\
AE (Spatial) \citep{Baur2020}                                                 & 0.313                   & 0.251             & 0.355                  & 0.349            \\
AE (Spatial) + CT \citep{Baur2020}                                            & 0.354                   & 0.315             & 0.377                  & 0.325            \\
AE (SSIM) \citep{Bergmann2019}                                                    & 0.349                   & 0.315             & 0.355                  & 0.352            \\
AE (SSIM) + CT \citep{Bergmann2019}                                               & 0.346                   & 0.310             & 0.347                  & 0.328            \\
F-AnoGan \citep{Schlegl2017}                                                  & 0.366                   & 0.361             & 0.401                  & 0.384            \\
VAE (Dense) \citep{Baur2020}                                                  & 0.371                   & 0.381             & 0.422                  & 0.392            \\
VAE (Dense) + CT \citep{Baur2020}                                             & 0.351                   & 0.342             & 0.419                  & 0.396            \\
VQ-GAN + Transformer (Codebook optimised 3D GAN variant of \citep{Pinaya2021}) & 0.463                   & 0.410             & 0.509                  & 0.453            \\
VQ-GAN + Transformer + CT conditioning (ours)                                                  & 0.497                   & 0.473             & 0.551                  & 0.521            \\
VQ-GAN + Transformer + CT conditioning + KDE (ours)                                            & 0.562                   & \textbf{0.561}             & 0.575                  & \textbf{0.579}            \\
VQ-GAN + Transformer + CT conditioning + KDE + 40\% Thresholding (ours)                        & \textbf{0.598}                   & 0.532             & \textbf{0.612}                  & 0.551   \\
\hline
\end{tabular}}
\end{table*}

\subsubsection{Ablation study:}
\begin{figure*}[h]
\centering
\includegraphics[width=\textwidth]{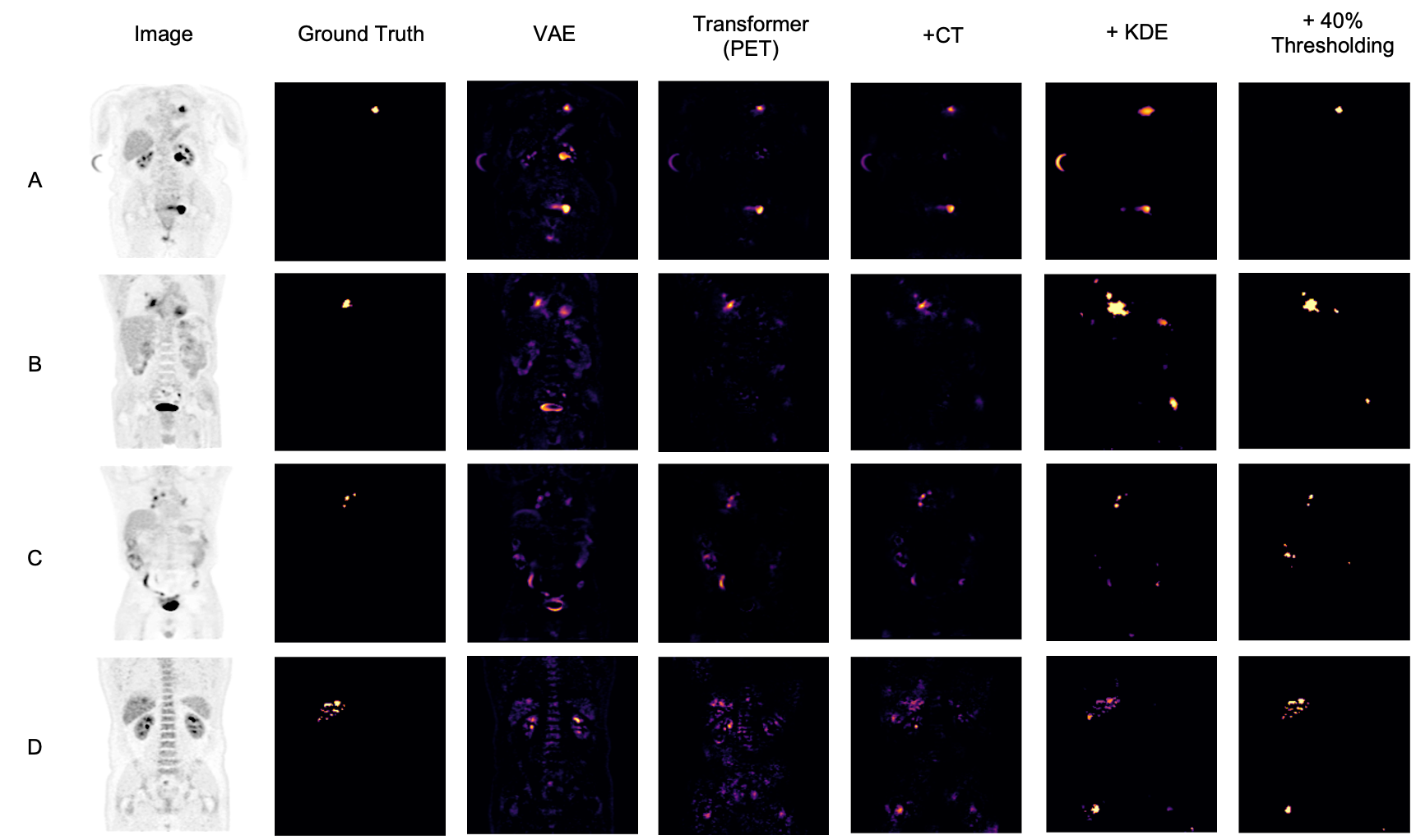}
\caption{Columns from left to right display (1st) the input image; (2nd) the gold standard truth segmentation; (3rd) the anomaly map as the residual for the PET only VAE, (4th) Transformer, and (5th) CT conditioned methods; (6th) the abnormality map as a KDE, (7th) and after thresholding at 40\% of each abnormal region maximum values. Results are provided for four randomly chosen subjects (A,B,C,D)} \label{testing_comparisons}
\end{figure*}

For the models trained on anomalous training data we observe a statistically significant improvement $(P < 0.005)$ in anomaly detection DICE performance by implementing CT conditioning compared to the 3D VQ-GAN variant approach of \cite{Pinaya2021}. A statistically significant improvement in AUCPR is also recorded $(P<0.001)$. This result confirms our initial expectations on the use case of anatomical context in the case of whole-body PET. Given the variability of healthy radiotracer uptake patterns, it is expected that beyond common areas like the bladder, further context is required to identify uptake as physiological or pathological. By incorporating model uncertainty to generate KDE maps, we see a further improvement in the overall DICE score and an even greater increase in AUPRC from 0.473 to 0.561 against the CT conditioned model $(P < .001)$. This behaviour can be explained by the increased variability around heterogeneous areas of healthy uptake, attributing to a decrease in false positives. The main advantage, as visualised in Fig. \ref{testing_comparisons}, is the increase in precision. By discarding the assumption of Gaussian uptake distributions, the model can better differentiate patterns of physiological uptake from pathological whilst still being sensitive to subtle anomalies, as seen in sample C in Fig. \ref{testing_comparisons}. We also see the same improvements across models in DICE and AUPRC for the models trained on normal data.

\subsubsection{Comparison to state-of-the-art: }

From Table 2, we can see a statistically-significant improvement $(P < .001)$ presented via the VQ-GAN + transformer approach using only PET data in relation to all autoencoders with and withouth CT inputs, the variational autoencoder and F-AnoGan method. This result is expected, as demonstrated in prior research \citep{Pinaya2021}. However, this divergence is also attributed to the presence of anomalies during training. It can be observed in some samples that the autoencoder method performs worse on large anomalies as it is able to partially or fully reconstruct them. We can see looking at the autoencoder methods trained on healhty data that performance has improved, however even given this, our proposed method still outperforms all baselines. Comparing the method proposed by \cite{Pinaya2021} to our best model comprising of CT conditioning and KDE anomaly maps, our approach generates an improvement in DICE score from 0.463 to 0.562 $(P < .001)$ with a considerable increase in AUPRC from 0.410 to 0.561 $(P < .001)$. Finally, through clinically accurate segmentations by growing segmented regions, we see a large increase in the best possible DICE score, but a reduction in AUPRC, likely brought about by the expansion of false-positive regions.

\subsection{Golden Out-of-sample Testing}
As a further test of the proposed models and their ability to generalize to unseen data from a different source/dataset, we showcase the performance of each proposed model and state-of-the-art comparisons on fully out-of-sample data using the CPTAC dataset. Pictorial results can be seen in Fig. \ref{OOS_comparisons}, with numerical performance visible in table \ref{oos_testing_results}. Note that the performance on the golden hold-out sample was, surprisingly, found to be better than in the previous experiments, suggesting that Dice scores are highly dependent on the type and size of the cancers. 

\begin{table*}[h]
\centering
\caption{Anomaly detection results on out-of-sample testing data. The performance is measured with best achievable DICE-score ($\lceil DICE \rceil$) and AUPRC on the test set.}\label{oos_testing_results}
\resizebox{15cm}{!}{%
\begin{tabular}{lcc|cc}
\hline
\textbf{Method}                                                                                & \multicolumn{2}{c|}{Anomalous Training Data} & \multicolumn{2}{c}{Normal Training Data} \\
\hline
& $\lceil DICE \rceil$              & AUPRC             & $\lceil DICE \rceil$             & AUPRC          \\        \hline
AE (Dense) \citep{Baur2020}                                                   & 0.359                & 0.344                & 0.438               & 0.386            \\
AE (Dense) + CT \citep{Baur2020}                                              & 0.437                & 0.403                & 0.423               & 0.374            \\
AE (Spatial) \citep{Baur2020}                                                 & 0.351                & 0.338                & 0.382               & 0.348            \\
AE (Spatial) + CT \citep{Baur2020}                                            & 0.358                & 0.342                & 0.385               & 0.349            \\
AE (SSIM) \citep{Bergmann2019}                                                     & 0.371                & 0.374                & 0.376               & 0.379             \\
AE (SSIM) + CT \citep{Bergmann2019}                                               & 0.365                & 0.373                & 0.364               & 0.371            \\
F-AnoGan \citep{Schlegl2017}                                                  & 0.424                & 0.408                & 0.415               & 0.375            \\
VAE (Dense) \citep{Baur2020}                                                 & 0.402                & 0.371                & 0.436               & 0.413            \\
VAE (Dense) + CT \citep{Baur2020}                                            & 0.394                & 0.369                & 0.447               & 0.408            \\
VQ-GAN + Transformer (Codebook optimised 3D GAN variant of \citep{Pinaya2021}) & 0.453                & 0.385                & 0.471                   & 0 .405           \\
VQ-GAN + Transformer + CT conditioning (ours)                                                   & 0.500                & 0.468                & 0.529                   & 0.511             \\
VQ-GAN + Transformer + CT conditioning + KDE (ours)                                            & 0.610                & 0.604                & 0.618                   & 0.600             \\
VQ-GAN + Transformer + CT conditioning + KDE + 40\% Thresholding (ours)                        & \textbf{0.717}       & \textbf{0.631}       & \textbf{0.719}         & \textbf{0.642}   \\
\hline
\end{tabular}}
\end{table*}

\begin{figure*}[!t]
\centering
\includegraphics[width=10cm]{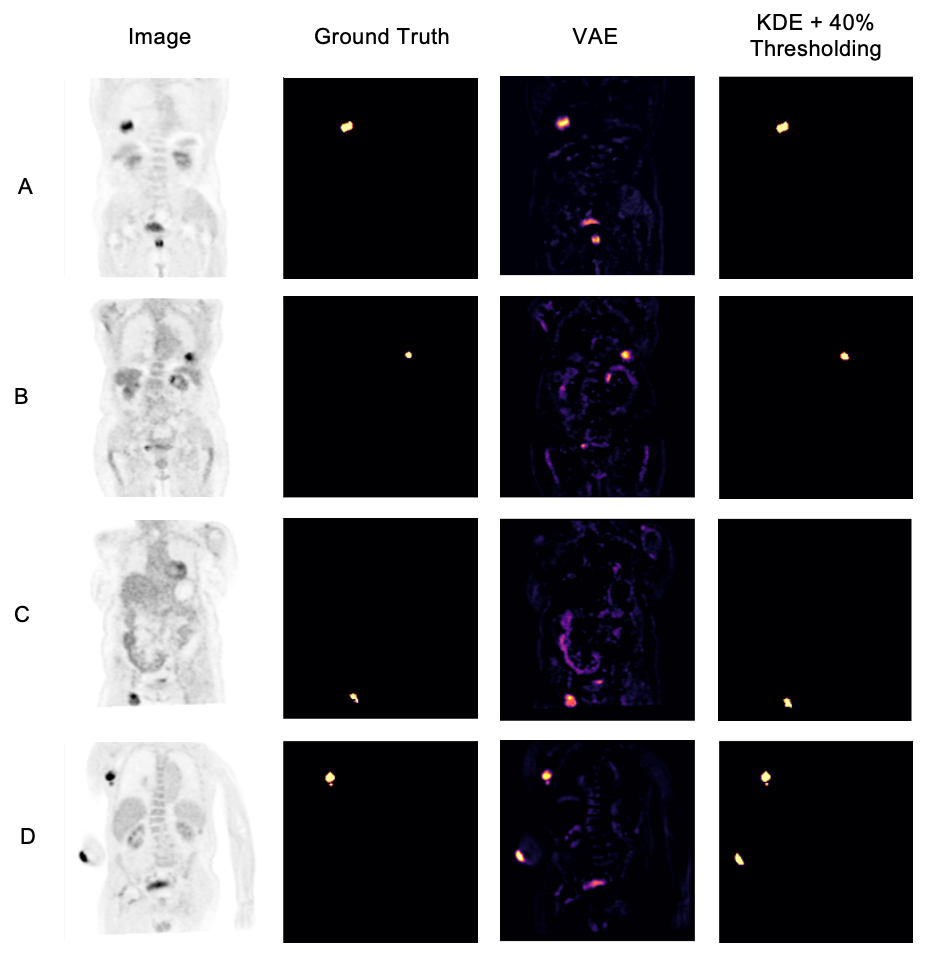}
\caption{Columns from left to right display (1st) the input image; (2nd) the gold standard truth segmentation; (3rd) the anomaly map as the residual for the VAE, (4th) our approach using the abnormality map as a KDE and after thresholding at 40\% of each abnormal region maximum values. Results are provided for four randomly chosen subjects (A,B,C,D)} \label{OOS_comparisons}
\end{figure*}

\section{Discussion}
From the results, there is strong evidence and motivation for using multi-modal conditioning for whole-body PET anomaly detection and using a KDE approach for producing anomaly maps. Not only do the proposed methods generate improved results over current state-of-the-art, but their performance is able to generalise to out-of-sample data and perform to the same level of competency. Firstly, we demonstrate the impact of appropriate codebook sizes when employing discrete latent space based representations for anomaly detection with autoregressive models. This alone showed a large range in performance based on changes using PET data only; additionally, this impact was further shown with respect to the size of the codebook for the conditioning CT data. Furthermore, with respect to the KDE approach for generating anomaly maps, we showcase the importance of an appropriate kernel shape and bandwidth regularisation term, which can generate large performance improvements when optimised properly. We should note however that the efficacy of CT conditioning is greatly brought about through the attention mechanism used in transformer that allows the model to effectively model the relationship between the PET and CT data. We can see in the baseline methods that when both PET and CT data is used, there is little to no performance increase. Additionally we showcase how our approach works well on models trained on both normal and anomalous training data. We can see that there is a slight improvement when trained on normal data, as to be expected, however in the field of medical imaging where normal data can often be hard to come by, being able to produce a model via unsupervised approaches whilst still generating comparable results is a key breakthrough of this work.
There are, however, still areas for improvement beyond the current scope of this research. We still see varying cases of false positives across samples, showing ongoing difficulties in differentiating physiological from pathological uptake. The reasons for this are likely patient-specific and can be down to several factors, i.e., pre-existing diseases, general health, age, PET/CT miss-alignment, or whether the scan was performed in fasting state or not. A further example can be seen in sample A, Fig \ref{testing_comparisons}, where the injection site can be visualised in the patient’s arm (although traditionally PET scans are performed in the “arms up” position). Additionally, we see in Sample D in Fig. \ref{OOS_comparisons} that a patient with an amputated arm showcases high uptake at the amputation site recorded as a false positive. Naturally, one solution would be to provide more training data increasing observed variability; however, further work to combat these issues could stem from some form of weak or self-supervision via to help to remove the outliers from the training data. Another approach could also be to incorporate some form of ensembling of models at different levels of downsampling within the VQ-GAN. In doing so the combined models will showcase greater local and global context such that the model might be able to better differentiate healthy physiological uptake from pathological uptake.

\section{Conclusion}
Detection and segmentation of anomalous regions, particularly for cancer patients, is essential for staging, treatment and intervention planning. In this study, we propose a novel  transformer-based anomaly detection model using multi-modal conditioning and kernel density estimation via model stochasticity. Proposed model achieves statistically-significant improvements in Dice and AUPRC, representing a new state-of-the-art compared to competing methods. We further show the impact of codebook size selection to act as a key consideration when implementing VQ-VAE based methods. In addition, we show that the kernel choice and bandwidth regularisation for the kernel density estimation approach significantly impact the anomaly detection performance when using KDE anomaly maps, a superior alternative to residual maps. We show the impact of proposed methods when faced only with training data containing anomalies, showing greater robustness than autoencoder-only approaches. The strong evidence presented here indicates that multi-modal abnormality detection models, when combined with the proposed KDEs method, are key features that deserve further focus and development by the community. 

\acks{This research was supported by Wellcome/ EPSRC Centre for Medical Engineering 
\\
(WT203148/Z/16/Z), Wellcome Flagship Programme 
(WT213038/Z/18/Z), The London AI Centre for Value-based Heathcare and GE Healthcare. The models were trained on the NVIDIA Cambridge-1, UK’s largest supercomputer, aimed at accelerating digital biology.}

%
\ethics{The work follows appropriate ethical standards in conducting research and writing the manuscript, following all applicable laws and regulations regarding treatment of animals or human subjects.}

\coi{We declare we don't have conflicts of interest.}

\bibliography{journal_06_22}


\clearpage
\appendix
\section{Model Parameters}
\label{model_params}

The total number of parameters for the VQ-GAN Model used to encode the PET and CT along with the parameters of the Performer models used in this work can be outlined in Table \ref{tab:num_params}.

\begin{table}[h]
\centering
\caption{Number of parameters of models used}\label{tab:num_params}
\begin{tabular}{ccc}
\hline
Model     & Trainable Parameters & Total Parameters \\
\hline
VQ-GAN             & 24,599,681           & 24,632,449       \\
Performer PET only & 18,684,673           & 18,684,673       \\
Performer PET + Cross attention CT & 33,396,993           & 33,396,993  \\
\hline
\end{tabular}
\end{table}

\section{Model AUC Curves}
\label{auc_curves}

Furthermore we display a number of meaningful Precision Recall curves for selected models on both testing datasets to better visualise the sensitivity and specificity tradeoff of the proposed methods.

\begin{figure*}[h]
\centering
\includegraphics[width=10cm]{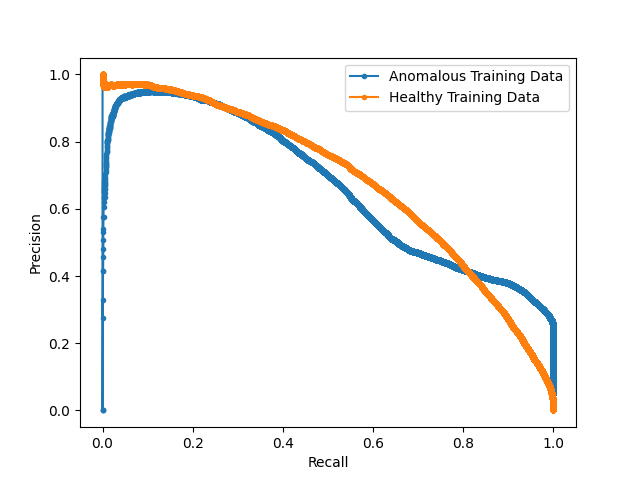}
\caption{Precision recall curves for proposed model using PET and CT data with KDE anomaly maps trained on anomalous and healthy training data, for testing outlined in section \ref{Testing1}} \label{OOS_comparisons}
\end{figure*}

\begin{figure*}[h]
\centering
\includegraphics[width=10cm]{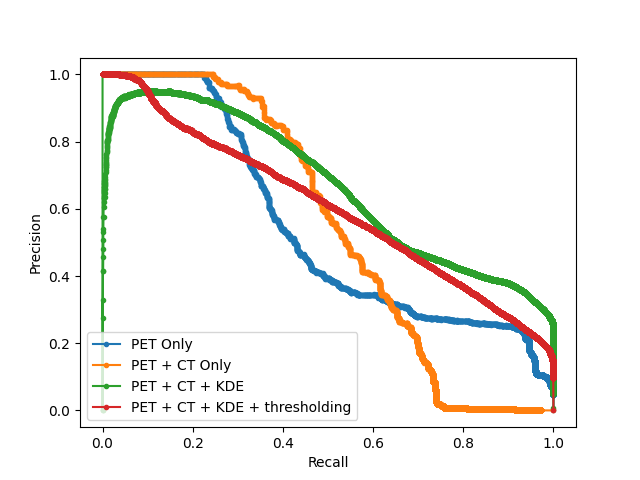}
\caption{Precision recall curves for ablation study of VQ-VAE + Transformer along with additions including CT conditioning, KDE anomaly maps and thresholding. Results presented for testing outlined in section \ref{Testing1}} \label{OOS_comparisons}
\end{figure*}

\begin{figure*}[h]
\centering
\includegraphics[width=10cm]{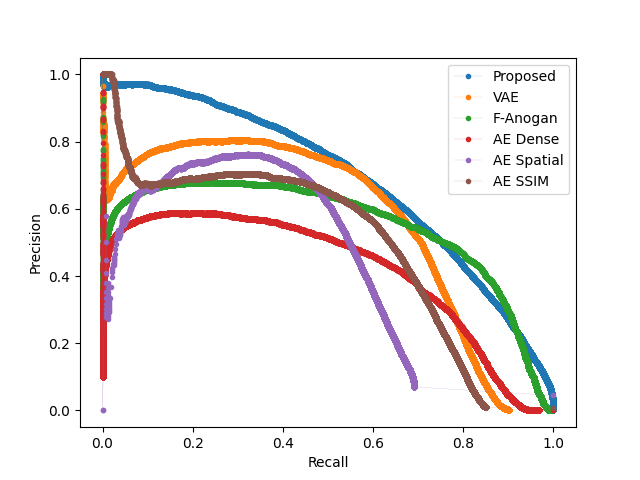}
\caption{Precision recall curves comparing baseline methods against our proposed method of the VQ-VAE + Transformer with CT conditioning and KDE anomaly map trained on healthy data. Results presented for testing outlined in section \ref{Testing1}} \label{OOS_comparisons}
\end{figure*}

\begin{figure*}[t!]
\centering
\includegraphics[width=10cm]{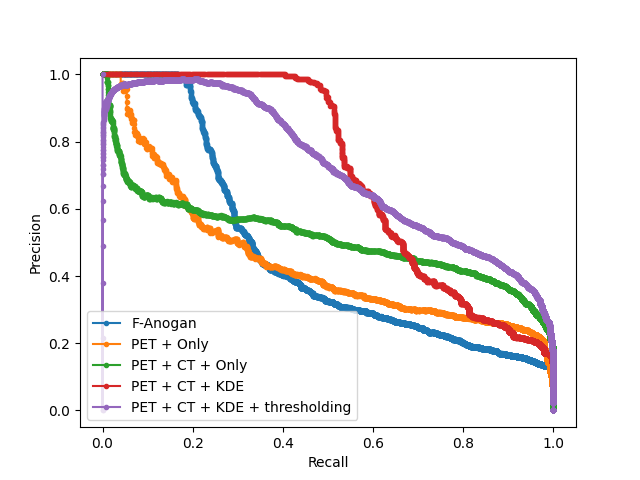}
\caption{Precision recall curves for ablation study of VQ-VAE + Transformer along with additions including CT conditioning, KDE anomaly maps and thresholding. Additionally we showcase the performance of the best alternative baseline model for the CPTAC testing set - F-AnoGan. Results presented for testing on the out-of-sample testing on CPTAC for models trained on anomlaous data.} \label{OOS_comparisons}
\end{figure*}

\end{document}